\documentclass{nature}

\usepackage{amssymb}
\usepackage{changepage}
\usepackage{enumitem}
\usepackage{graphicx}
\let\realincludegraphics\includegraphics
\AtBeginDocument{\let\includegraphics\realincludegraphics}
\usepackage{array}
\usepackage{hyperref}
\newcolumntype{C}[1]{>{\centering\arraybackslash}p{#1}}
\usepackage{multirow}
\usepackage[capitalize]{cleveref}

\usepackage[font=small]{caption}

\title{Depressed individuals express more distorted thinking on social media.}

\author{Krishna C. Bathina$^{1}$, Marijn ten Thij$^1$, \& Lorenzo Lorenzo-Luaces$^2$, Lauren A. Rutter$^2$, Johan Bollen$^{1}$}

\begin{document}
\maketitle

\begin{enumerate}[itemsep=-1ex,partopsep=1ex,parsep=1ex]
\item Center for Social and Biomedical Complexity, Indiana University
\item Department of Psychological and Brain Sciences, Indiana University.
\end{enumerate}

\begin{abstract}
Depression is a leading cause of disability worldwide, but is often under-diagnosed and under-treated. One of the tenets of cognitive-behavioral therapy (CBT) is that individuals who are depressed exhibit distorted modes of thinking, so-called cognitive distortions, which can negatively affect their emotions and motivation. Here, we show that individuals with a self-reported diagnosis of depression on social media express higher levels of distorted thinking than a random sample. Some types of distorted thinking were found to be more than twice as prevalent in our depressed cohort, in particular  \emph{Personalizing} and \emph{Emotional Reasoning}. This effect is specific to the distorted content of the expression and can not be explained by the presence of specific topics, sentiment, or first-person pronouns. Our results point towards the detection, and possibly mitigation, of patterns of online language that are generally deemed depressogenic. They may also provide insight into recent observations that social media usage can have a negative impact on mental health.
\end{abstract}

\section{Introduction}
Depression is a leading contributor to the burden of disability worldwide\cite{greenberg2015economic,WHO}, with some evidence that disability attributed to depression is rising, particularly among youth\cite{case2015rising,mojtabai2016national}. A key challenge in reducing the prevalence of depression has been that it is often under-recognized\cite{mitchell2009clinical} as well as under-treated\cite{wang2005twelve}. Cognitive-behavioral therapy (CBT), is the most widely researched psychotherapy for depression. It is equivalent to antidepressant medications in its short-term efficacy and evidences superior outcomes in the long-term\cite{hofmann2012efficacy,cuijpers2013does}. The cognitive theory underlying CBT argues that the ways in which individuals process and interpret information about themselves and their world is directly related to the onset, maintenance, and recurrence of their depression\cite{beck2014advances, clark2010cognitive}. This model is consistent with information processing accounts of mood regulation\cite{foland2012cognitive} and its dynamics\cite{leemput:critical2014}, as well as basic research that supports the role of cognitive reappraisal and language in emotion regulation\cite{webb2012dealing, bollen2019, derubeis2008cognitive, troy2010seeing}. 

In CBT, therapists work with their clients to identify depressogenic thinking patterns by identifying lexical or verbal markers of rigid, distorted, or overly negative interpretations\cite{lorenzo2015s, ozdel2014measuring}. For example, statements that include ``should'' or ``must'' are often challenged as reflecting overly rigid rules about the world (``I shouldn't be lazy'', ``I must never fail''). This process often entails a series of conversations with the client to uncover and address statements that reflect these so-called Cogntive Distortions (CD).

The idea that language is predictive of depression is supported by data-driven approaches detecting depression from various lexical markers including the use of language to describe negative emotions\cite{rude2004language,tackman:depression2019}, the use of first-person pronouns\cite{bernard2016depression,smirnova:language2018,zimmermann2017first,cacheda:early2019}, and mentions of common symptoms\cite{content:cavazos2017}. Machine learning approaches have been shown to successfully predict whether Facebook users suffer from depression\cite{Eichstaedt11203,predict:dechoudhury2013}, identifying the most useful lexical markers to render a prediction. These results, while useful for prediction and the detection of depression, do not offer insights into the cognitive dynamics of the disease pattern, nor its relationship to language, which is crucial in developing treatments and interventions.

Here, we emphloy a theory-driven approach to studying depressive language on Twitter. Rather than attemphting to extract relevant text features from text data, e.g.~``sleep'', ``health'', or other mental health related features, we define a clinical lexicon of 241 n-grams\cite{jurafsky:speech2008} that a panel of clinical psychologists deemed to form a schema involved in the expression of a particular type of distorted thinking according to CBT theory and practice. For example, ``I will never \_'' would be implicated in the expression of a cognitive distortions such as \emph{Catastrophizing} or \emph{Fortune-telling}, whereas ``I am a \_'' would be used to express a \emph{Labeling and Mislabeling} distortion.

We then compare the longitudinal prevalence of this set of Cognitive Distortion Schemata (CDS) in the language of a large cohort of depressed individuals vs.~a random sample on social media (Twitter). Our results indicate significantly higher prevalence of most types of CDS in the Depressed cohort, both at the within-subjects and between-groups level. Particularly CDS in the \emph{Personalizing} and \emph{Emotional Reasoning} types occur approximately 2.3 times more frequently in the online language of Depressed users. Our results are robust to changes in our user sample, our choice of CDS n-grams, text sentiment, and the known propensity of Depressed individuals to make self-referential statements.

\subsection{Cognitive distortion types and n-gram schemata}

Aaron T. Beck introduced the concept of cognitive distortions to characterize the thinking of individuals with depression\cite{beck1963thinking,beck1964thinking}. Subsequently, other clinicians expanded on his typology of distortions\cite{burns1989}, including most recently his daughter, clinical psychologist and CBT expert, Judith Beck\cite{beck1995cognitive}. We drew upon these latest lists to identify 12 types of cognitive distortions that may characterize the thinking of individuals who are depressed. 

\begin{table*}[!ht]
\scriptsize
\caption*{Cognitive Distortion Types}
\begin{center}
\begin{tabular}{p{4.5cm}p{6cm}p{5cm}}
Category                            &   Definition          &       Examples        \\ \hline

Catastrophizing            &   Exaggerating the importance of negative events
                                    &   "\textit{The evening \textbf{will be a disaster}}"\\

Dichotomous Reasoning      &   Thinking that an inherently continuous situation can only fall into two categories 
                                    &   ``\textit{\textbf{No one} will \textbf{ever} like me.}''\\
                                    
Disqualifying the Positive &   Unreasonably discounting positive experiences  
                                    &   ``\textit{\textbf{OK but}$^1$ my grade \textbf{was not that good.$^2$}}'' \\
                                    
Emotional Reasoning        &   Thinking that something is true based on how one feels, ignoring the evidence to the contrary
                                    &   ``\textit{My grades are good but it \textbf{still feels}$^1$ like I \textbf{will fail}$^2$.}''\\
                                    
Fortune-telling            &   Making predictions, usually negative ones, about the future.
                                    &   ``\textit{Whatever I try \textbf{I will not} be successful}''\\

Labeling and Mislabeling   &   Labeling yourself or others while discounting evidence that could lead to less disastrous conclusions
                                    &   ``\textit{\textbf{I am a}$^1$ \textbf{total}$^2$ \textbf{loser}$^3$.}''\\

Magnification and Minimization &   Magnifying negative aspects or minimizing positive aspects
                                        &   ``\textit{My good grades are really \textbf{not important}.}'' \\
                                        
Mental Filtering           &   Paying too much attention to negative details instead of the whole picture
                                    &   ``\textit{\textbf{If I only} worked harder, I would be more successful.}''\\
                                    
Mindreading                &   Believing you know what others are thinking
                                    &   ``\textit{\textbf{Everyone believes}$^1$ \textbf{I am a}$^2$ \textbf{failure}$^3$.}''\\
                                    
Overgeneralizing           &   Making sweeping negative conclusions based on a few examples
                                    &   ``\textit{\textbf{Nobody ever} cares for me.}''\\
                                    
Personalizing              &   Believing others are behaving negatively because of oneself, without considering more plausible or external explanations for behavior
                                    &   ``\textit{\textbf{Everyone thinks}$^1$ \textbf{I am a loser}$^2$ for calling her.}''\\
                                    
Should Statements          &   Having a fixed idea on how you and/or others should behave
                                    &   ``\textit{I \textbf{have to}$^1$ to do this or \textbf{I will not}$^2$ make it to the weekend.}''\\
                                    
\end{tabular}
\end{center}
\caption{Common cognitive distortions identified in CBT for depression\cite{beck1995cognitive}. We provide examples of how Cognitive Distortion Schemata (CDS) can be embedded in common expressions (bold). Contractions are expanded. Numbers indicate where the example contains more than one CDS.}
\label{table:CD_definitions}
\end{table*}

We defined 241 CDS n-grams in total, each expressing at least 1 type of cognitive distortion (see Appendix Table 7). The schemata in each category were formulated to capture the ``minimal semantic building blocks'' of expressing distorted thinking for the particular type, avoiding expressions that are specific to a depression-related topics, such as poor sleep or health issues. For example, the 3-gram ``I am a'' was included as a building block of expressing \emph{Labeling and Mislabeling}, because it would be a highly likely (and nearly unavoidable) n-gram to express many self-referential (``I'') expressions of labeling (``am a'') (for an example see \cref{table:CD_definitions}). Where possible, higher-order n-grams were chosen to capture as much of the semantic structure of one or more distorted schemata as possible, e.g. the 3-gram ``everyone will believe'' captures both \emph{Overgeneralizing} and \emph{Mindreading}. We did include 1-grams such as ``nobody'' and ``everybody'' in spite of their prevalence in common language, since they strongly correspond to the expression of \emph{Dichotomous Reasoning}. \cref{table:CD_classes} shows the number of schemata per category in our CDS set along with the average n-gram size, and a number of relevant grammatical features. The complete set of CD schemata is provided in Table 7 in the Appendix.

We note that a significant sub-set of the CDS do not occur in the Twitter content for both cohorts (see \cref{table:CD_classes}: $N_\exists$), indicating that parts of our set of CDS are ``lexically exhaustive'' with respect to capturing the major modes of CD expression in natural language.

\begin{table*}
\scriptsize
\begin{center}
\begin{tabular}{p{5.5cm}p{0.5cm}p{0.0cm}p{0.5cm}p{0.75cm}p{1.3cm}p{0.0cm}C{1.5cm}C{1.5cm}}
&                               & &   \multicolumn{3}{c}{Significant $N$} & & Avg. Length  & Pronouns \\\cline{4-6}\cline{8-9}
CD Category                     & $N_{CD}$     & & $N_\exists$  & $N*$   &$N^*_r$ (\%)  & & $\bar{n}$   &$P_r$ (\%)        \\ \hline
Catastrophizing                 & 11    & & 10           & 2      & 18.2    & & 3.000       & $/$         \\ 
Dichotomous Reasoning           & 23    & & 23           & 16     & 69.6   & & 1.347       & $/$         \\
Disqualifying the Positive      & 14    & & 13           & 4      & 28.6   & & 2.286       & 7.1       \\ 
Emotional Reasoning             & 7     & & 7            & 6      & 85.7   & & 2.857       & 85.7       \\ 
Fortune-telling                 & 8     & & 8            & 8      & 100.0   & & 3.125       & 87.5       \\ 
Labeling and Mislabeling        & 44    & & 44           & 20     & 45.5   & & 2.273       & 36.4       \\ 
Magnification and Minimization  & 8     & & 8            & 4      & 50.0   & & 2.000       & $/$         \\ 
Mental Filtering                & 14    & & 14           & 3      & 21.4   & & 2.786       & 50.0       \\ 
Mindreading                     & 72    & & 62           & 9     & 12.5   & & 3.167       & 83.3       \\ 
Overgeneralizing                & 21    & & 21           & 14     & 66.7   & & 2.762       & 57.1       \\ 
Personalizing                   & 14    & & 14           & 9      & 64.3   & & 2.429       & 100.0       \\ 
Should Statements               & 5     & & 5            & 2      & 40.0   & & 1.400       & $/$         \\ \hline
Total                           & 241   & & 229          & 97    & 40.2   & & 2.585       & 51.0       \\
\end{tabular}
\end{center}
\caption{Descriptive statistics of our set of Cognitive Distortion Schemata which are grouped in 12 types (``CD Category''). Column $N_{CD}$ indicates the number of schemata in the specific category. \emph{Mindreading} is the largest category because it contains many grammatical variations of the same schema, e.g. ``I\slash you\slash they \_ will think''. The $N_\exists$ column shows the number of n-grams in each category that were actually found in our Twitter data. The $N*$ and  $N^*_r$ columns show respectively the number and ratio of n-grams in the category for which we found a statistically significantly higher prevalence in the Depressed than the Random Sample cohort. $\bar{n}$ is the average length (n) of each n-gram in the category. Column $P_r$ shows the ratio of CDS in each category that contain a first-person pronoun. $/$ indicates the absence of any CDS with the feature. }
\label{table:CD_classes}
\end{table*}

\subsection{Depressed and random sample}
We identified a cohort of social media users that had self-reported a clinical diagnosis of depression by posting a variant of the explicit statement ``I was diagnosed with depression'' (see ``Materials and Methods''). To make sure we were only including truly self-referential statements of diagnosis of depression, 3 of the authors manually removed quotes, retweets, jokes, and external references. Note that we exclude all diagnosis statements themselves from our analysis, including all tweets that contain the term ``diagnos'' and ``depress''. We also examine the sensitivity of our results to the propensity of this cohort to make similar self-referential statements (see ``Absence of personal pronoun effect.'') 

With this final set of adjudicated diagnosis tweets, we harvested the maximum number of tweets allowed by the Twitter API (the most recent 3200) for each individual, resulting in a sample of 1,207 users and their 1,759,644 tweets (ranging from May 2008 to September 2018). We refer to this cohort as ``Depressed'', but acknowledge that we have no independent confirmation of their present mental health state. We also established a baseline sample of randomly chosen individuals with a similar distribution of account creation dates as the Depressed cohort to account for changes in user behavior and platform effects. Here too we exclude all tweets that contain the terms ``diagnos'' and ``depress'' from subsequent analysis. Our ``Random Sample'' cohort contains 8,791 individuals and a total 8,498,574 tweets (see ``Materials and Methods'').

\section{Results}

\begin{figure}[!htbp]
\centering
\includegraphics[width=0.7\linewidth]{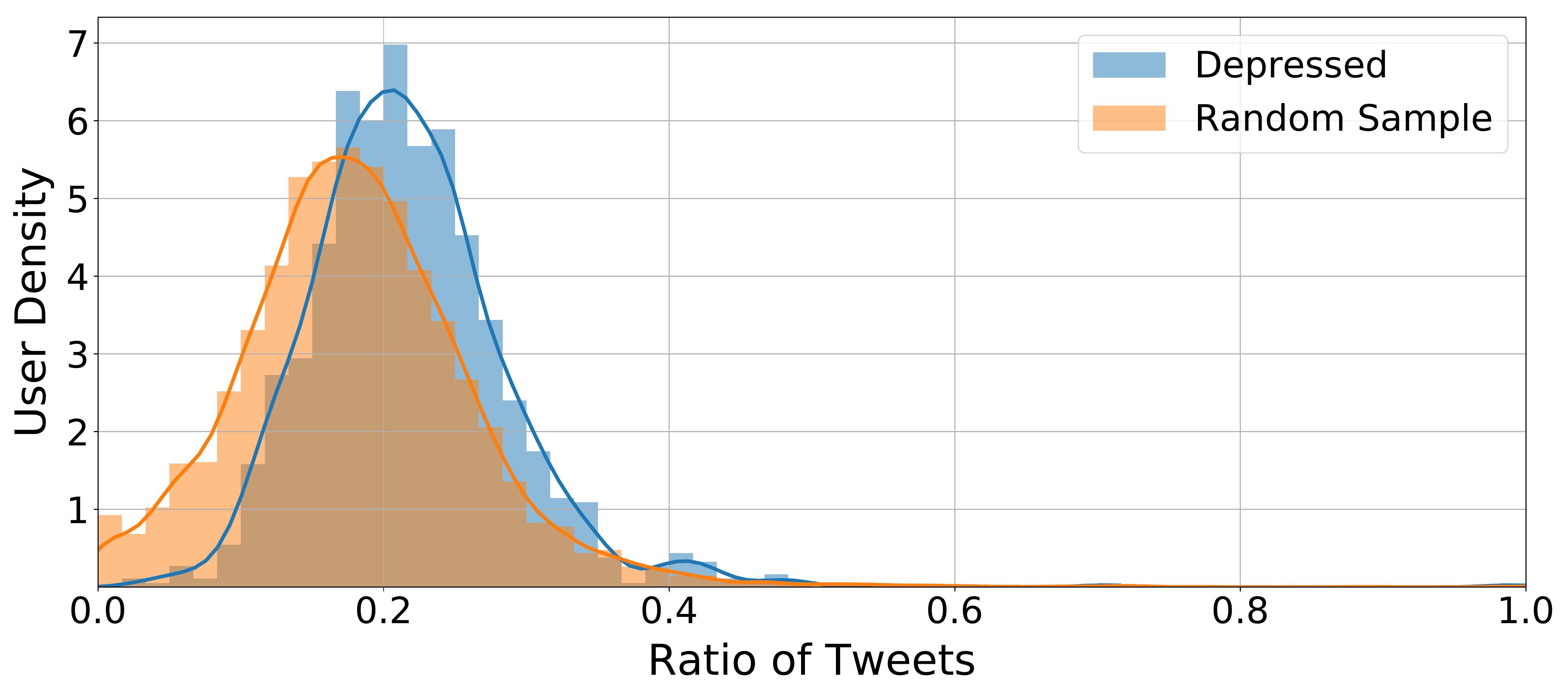}
\caption{
Density of \textit{within-subject} prevalence of tweets containing CDS for individuals with at least 150 tweets shows a greater number of individuals with high CDS levels in the Depressed than in the Random Sample cohort.
The results of a two-sample K–S test allows us to reject the null-hypothesis that the two distributions are drawn from the same sample ($N_d = 1100$  (depressed), $N_r = 6151$ (random), $p=1.036\times 10^{-45}$).}
\label{fig:User Ratio}
\end{figure}

We first compare the within-subject prevalence of the established set of CDS between the Depressed and Random Sample cohorts. For each individual we count how many of their tweets contained any of the 241 CDS and divide it by their total number of tweets, resulting in an individual \textit{within-subject} CD prevalence (see ``Materials and Methods''). The density distribution of individual prevalence values can then be compared between Depressed and Random Sample individuals as shown in \cref{fig:User Ratio}. We restrict this analysis to individuals with at least 150 tweets so that we have sufficient data to determine prevalence reliably, but retain all individuals in subsequent between-groups analyses since the latter does not require the calculation of within-subject prevalence values.

We observe that the distribution of within-subject CDS prevalence is shifted significantly to the right for the Depressed cohort relative to that of the Random Sample, indicating that individuals in the Depressed cohort express significantly more CDS. Note that $0.487$\% of the Random Sample individuals have \textit{no} tweets with CDS whereas the Depressed sample has no cases without CDS. Results from a Two-Sample Kolmogorov–Smirnov test ($p < 0.0001$) indicate that we can reject the null hypothesis that the two samples are drawn from the same distribution.

\begin{figure}[!htbp]
\centering
\includegraphics[width=0.7\linewidth]{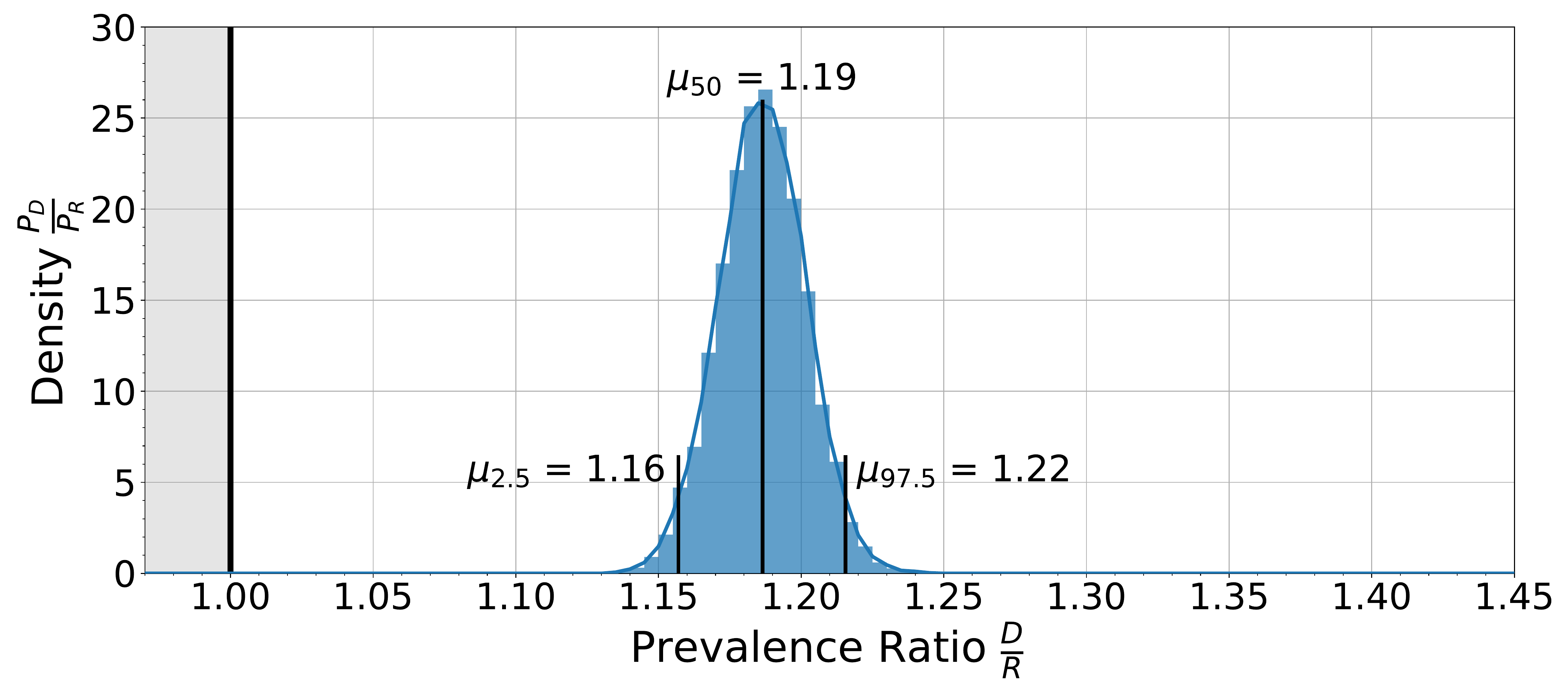}
\caption{
Density of bootstrapped \textit{between-groups} Prevalence Ratios, with median ($\mu_{50}=1.19$) and  95\% Confidence Intervals (CI) $[1.16,1.22]$ between the Depressed and Random Sample cohort (see \cref{table:CD_Category_Prevalence}: ``All CDs''). The 95\% CI of the distribution does not include $1.00$ (vertical line) indicating a significantly higher prevalence of all CDS for the Depressed cohort ($1.2\times$).}
\label{fig:All CD Categories}
\end{figure}

Furthermore, we conduct a between-groups analysis to compare the prevalence of CDS between the Depressed vs.~the Random Sample cohort. We do so by calculating the Prevalence of CDS for all tweets from each cohort and calculating the Prevalence Ratio ($PR$) between the two cohorts (see Materials and Methods ``Prevalence Ratio''). A Prevalence Ratio significantly larger than 1 indicates that the presence of CDS in the tweets written by the Depressed cohort is greater than the Random Sample cohort. To assess the sensitivity of our results to changes in our cohort samples, we repeatedly calculate the estimated $PR$ value over 10,000 random re-samples (with replacement) of both groups, resulting in a distribution of PR values shown in \cref{fig:All CD Categories} (see Materials and Methods ``Bootstrapping''). Note, Prevalence Ratios express the \textit{relative} difference between the 2 cohorts, not the absolute difference which is provided in Appendix Table 6.

We observe in \cref{fig:All CD Categories} that the median of this distribution of PR values is significantly larger than 1 (and its 95\% confidence intervals does not include 1), indicating that we find a statistically significant higher prevalence of CDS in the Depressed cohort ($1.2\times$) than in the Random Sample, and that this result is robust to random changes in our cohorts.

\begin{table*}[!htbp]
\begin{center}
\scriptsize
\begin{tabular}{p{1.1cm}p{4.5cm}p{0.5cm}C{2cm}|p{0.5cm}C{2cm}|p{0.5cm}C{2cm}}
& &  \multicolumn{2}{c}{$PR_A$} 
& \multicolumn{2}{c}{$PR_1$} & \multicolumn{2}{c}{$PR_C$} \\
& & median & 95\% CI & median & 95\% CI & median & 95\% CI \\\cline{2-8}
\multirow{11}{*}{$P_D \gg P_R$}    &  
All CDS                                             & 1.186$^{*}$  & [1.157, 1.216] & 1.169$^{*}$ & [1.140, 1.197] & 1.220$^{*}$ & [1.160, 1.310] \\ \cline{2-8}
& Personalizing                                   & 2.402$^{*}$  & [2.242, 2.576] & \slash      & \slash         & 2.412$^{*}$ & [1.671, 2.957]\\
& Emotional Reasoning                             & 2.323$^{*}$  & [2.049, 2.639] & 2.065$^{*}$ & [1.702, 2.485] & 2.317$^{*}$ & [2.012, 3.184]\\ 	
& Overgeneralizing                                & 1.580$^{*}$  & [1.501, 1.661] & 1.486$^{*}$ & [1.409, 1.566] & 1.574$^{*}$ & [1.369, 1.734]\\
& Mental Filtering                                & 1.468$^{*}$  & [1.291, 1.656] & 1.346$^{*}$ & [1.069, 1.660] & 1.470$^{*}$ & [1.173, 1.919]\\	
& Labeling and Mislabeling                        & 1.328$^{*}$  & [1.267, 1.391] & 1.204$^{*}$ & [1.144, 1.268] & 1.319$^{*}$ & [1.156, 1.478]\\
& Disqualifying the Positive                      & 1.349$^{*}$  & [1.210, 1.498] & 1.349$^{*}$ & [1.210, 1.498] & 1.346$^{*}$ & [1.176, 1.555]\\	
& Dichotomous Reasoning                           & 1.195$^{*}$  & [1.163, 1.226] & 1.195$^{*}$ & [1.163, 1.226] & 1.216$^{*}$ & [1.158, 1.303]\\
& Mindreading                                     & 1.136$^{*}$  & [1.060, 1.230] & 1.136$^{*}$ & [1.060, 1.229] & 1.129       & [0.808, 1.274]\\
& Should Statements                               & 1.103$^{*}$  & [1.050, 1.153] & 1.103$^{*}$ & [1.050, 1.153] & 1.100       & [0.836, 1.409]\\
& Magnification and Minimization                  & 1.075$^{*}$  & [1.023, 1.130] & 1.075$^{*}$ & [1.023, 1.130] & 1.078$^{*}$ & [1.016, 1.472]\\\cline{2-8}
\multirow{2}{*}{$P_D \ll P_R$} & Fortune-telling  & 0.954        & [0.837, 1.075] & 0.586       & [0.483, 0.698] & 0.944       & [0.501, 1.288]\\	
& Catastrophizing                                 & 0.729        & [0.554, 0.902] & 0.729       & [0.554, 0.902] & 0.718       & [0.654, 1.062]\\		
\end{tabular}
\end{center}
\caption{Prevalence Ratio (PR) and 95\% CIs of CDS between the Depressed and Random Sample cohort. PR values $\gg 1$ indicate a significantly higher prevalence in the Depressed sample (marked by $^*$). Values are calculated under 3 distinct conditions, labeled $PR_A$: values for the entire set of CDS, $PR_1$: values for CDS without first-person pronouns (``I'',``me'', ``my'', ``mine'', and ``myself''), and $PR_C$: values with a 95\% CI resulting from resampling the CDS, instead of our sample of individuals (see Materials and Methods). Appendix Table 6 provides absolute Prevalence Differences.}
\label{table:CD_Category_Prevalence}
\end{table*}

\begin{figure*}[!htbp]
\centering
\includegraphics[width=0.8\linewidth]{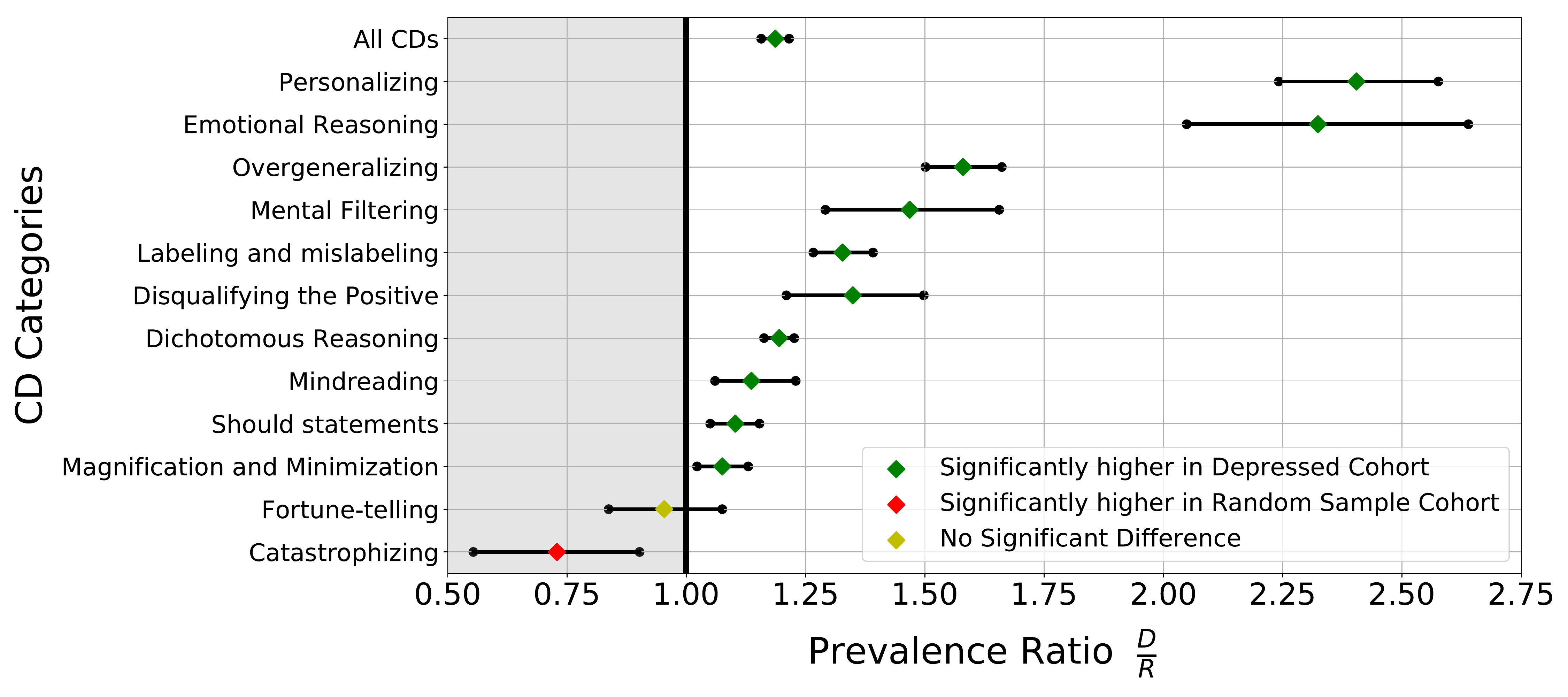}
\caption{Comparison of Cognitive Distortion Schemata (CDS) Prevalence Ratios between the Depressed and Random Sample cohort for each cognitive distortion type. Black dots indicate the boundaries of the 95\% CI. The medians are indicated by a green, yellow, or red diamond respectively corresponding to a higher prevalence in the Depressed, no difference, or higher prevalence among the Random Sample respectively. The Depressed cohort showed a significantly higher use of CDS than the Random Sample cohort for all CD types combined (``All CDS'') and most CDS types separately ($PR\gg1$) with the exception of \emph{Fortune-telling} and \emph{Catastrophizing}. Please see \cref{table:CD_Category_Prevalence} for more details about the Prevalence Ratios.}
\label{fig:Separated CD Categories}
\end{figure*}

The between-groups PR values shown in \cref{fig:All CD Categories} do not reflect specific distortion types; all CDS are equally and independently matched to all tweets. Total CDS prevalence over all tweets is 21.8\% and 18.407\% for the Depressed and Random Sample cohort respectively but differs significantly for each CD type (See Appendix Table 5). It is reasonable to expect that the different types of CDS may differ in their prevalence between our cohorts. We therefore repeat the above analysis, with CDS separated by CD type (see \cref{table:CD_classes}).

As shown in \cref{table:CD_Category_Prevalence} and \cref{fig:Separated CD Categories}, the prevalence of CDS is significantly higher for nearly all CD types in the tweets of the Depressed cohort than those of the Random Sample with Prevalence Ratio values ranging from $2.4\times$ to $1.1\times$, with the exception of \emph{Catastrophizing} and \emph{Fortune-telling}, with the latter not producing a PR significantly different from parity. The CD types \emph{Personalizing} and \emph{Emotional Reasoning} have the greatest PR values of $2.4\times$ and $2.3\times$, followed by \emph{Overgeneralizing} ($1.6\times$), \emph{Mental Filtering} ($1.5\times$), \emph{Labeling and Mislabeling} ($1.3\times$), and \emph{Disqualifying the positive} ($1.3\times$). The CD types \emph{Mind Reading}, \emph{Should Statements}, and \emph{Magnification and Minimization} have lower yet significant PR values of $1.1\times$. \cref{table:CD_classes} ``Significant N'' shows the number and ratios of schemata for each CD type that have PR values significantly different from parity. 

The PR individual CDS n-grams can differ significantly as well. Appendix Fig. 6 shows the contributions of each individual CDS n-gram separately. \cref{table:top10CDS} shows the CDS with the individually highest and lowest PR values to illustrate the CDS that are most prevalent in the Depressed and Random Sample cohort respectively. As shown, the highest ranked CDS for the Depressed cohort belong to the \emph{Mindreading}, \emph{Emotional Reasoning}, and \emph{Personalizing} type, whereas the highest ranked CDS for the Random Sample belong to the non-reflexive \emph{Mindreading} and \emph{Fortune-telling} type.

\begin{table}[!htbp]
\scriptsize
\begin{center}
\begin{tabular}{lcc}
PR rank         & Depressed             &   Random Sample   \\\hline
1               & everyone will think  &   we know \\
2               & since it feels        &   they will not  \\
3               & I caused              &   he believes  \\
4               & will go wrong         &   we believe \\
5               & because of my         &   we will not   \\
6               & because I feel        &   she will not  \\
7               & all my fault          &   an incompetent \\
8               & a burden              &   that will not  \\
9               & because my            &   we will know \\        
10              & I am always           &   we do not know \\ \hline
\end{tabular}
\end{center}
\caption{\label{table:top10CDS}Ten CDS with highest individual PR values between the Depressed and Random Sample cohorts to illustrate most prevalent n-grams in online language of either cohort.}
\end{table}

\subsection{Absence of sentiment effect}

Previous research has shown that the language of depressed individuals is less positive (lower text valence) and contains higher levels of self-referential language\cite{molendijk2010word,fast2010gender,al2018absolute,brockmeyer2015me,rude2004language,ingram1990self}. To determine the degree to which our results can be explained by text sentiment or self-referential statements instead of distorted thinking, we examine the valence loadings of our collection of tweets and CDS, and reproduce our results with and without CDS containing self-referential statements.

\begin{figure*}[!htbp]
\centering
\includegraphics[width=\linewidth]{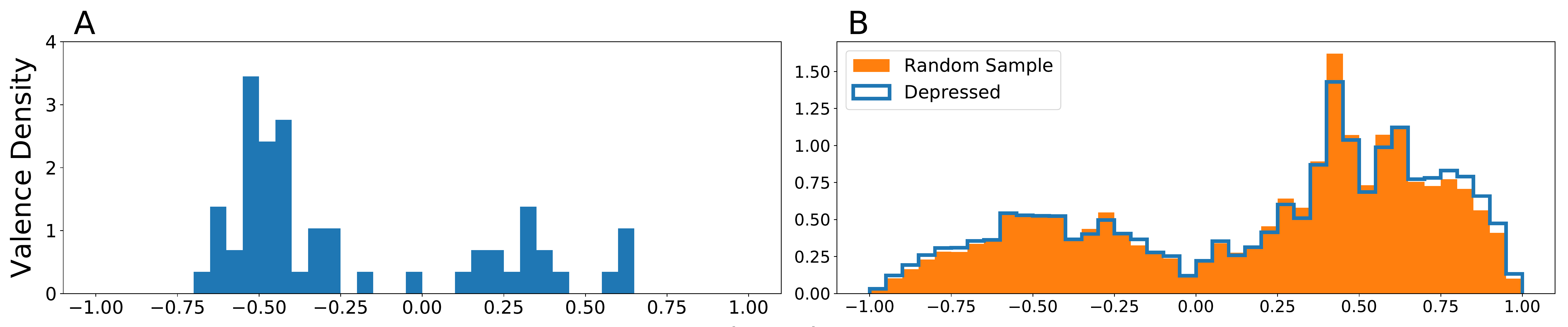}
\caption{(A) Density of Vader scores for CDS with non-zero sentiment values (24.1\% of 241 schemata). Most CDS carried no valence loading (75.9\%). The average rating for the complete set CDS is -0.05 (N=241). (B) Density of VADER valence ratings for all individual tweets posted by the Depressed and Random Sample cohorts, indicating a significant right-hand skew towards positive sentiment for both. The Depressed cohort has more extreme positive and negative sentiment than the Random Sample (blue line vs. orange bars). The results of a two-sample K–S test allows us to reject the null-hypothesis that the two sentiment distributions are drawn from the same distribution ($p<0.0001$).}
\label{fig:Vader Scores}
\end{figure*}

First, we determine the valence values of each CDS n-gram in our set using the VADER sentiment analysis tool\cite{hutto2014vader} which in a recent survey was shown to outperform other available sentiment analysis tools for social media language\cite{ribeiro2016}. VADER is particularly appropriate for this use, since its sentiment ratings take into account grammatical context, such as negation, hedging, and boosting. We find that 75.9\% of our CDS have either no sentiment-loaded content or are rated to have zero valence (neutral sentiment scores). The average valence rating of all CDS is $-0.05 (N=241)$ on a scale from $-1.0$ to $+1.0$. \cref{fig:Vader Scores}A shows the VADER sentiment distribution of only CDS n-grams with non-zero ratings. Here we observe only a slight negative skew of CDS sentiment for this small minority of CDS n-grams (24.1\%).

Furthermore, as shown in \cref{fig:Vader Scores}B, the sentiment distributions of all tweets for the Depressed and Random Sample cohorts are both skewed towards positive sentiment (right side of distribution). This matches earlier findings that human language exhibits a so-called Polyanna effect\cite{dodds:human2019}, a near-universal phenomenon that skews human language towards positive valence.  Surprisingly, we find no indications that the tweets of the Depressed cohort carry more negative valence than those of the Random Sample cohort. To the contrary, VADER sentiment ratings in the range $[0.70,1.00]$ seem to be slightly more prevalent among the tweets of the Depressed cohort (see \cref{fig:Vader Scores}B), possibly indicating an increased emotionality (higher levels of both negative and positive affect). One particular deviation in the sentiment range of $[0.40,0.45]$ was found to be uniquely associated with the Random Sample cohort using the ``Face With Tears of Joy'' emoji (VADER sentiment=0.4404) more often than the Depressed cohort. A two-sample K–S test allows us to reject the null-hypothesis that the two distributions are drawn from the same sample ($p<0.0001$\footnote{Value is below 32byte floating point precision.}).

Combined, these findings strongly suggest that the higher prevalence of CDS in the language of the Depressed cohort can neither be attributed to a negative valence skew in the CDS set, nor the sentiment distribution of the tweets produced by either the Depressed and Random Sample cohorts.

\subsection{Absence of personal pronoun effect}
Research has shown that First-Person Pronouns (FPP) are more prevalent in the language of depressed individuals\cite{rude2004language,smirnova:language2018}. Since many CDS contain FPPs (see \cref{table:CD_classes} ``Pronouns''), our results may to a degree reflect this phenomenon instead of the ``distorted'' nature of our CDS. To test the sensitivity of our results to the presence of FPPs in our set of CDS, we repeat our analysis entirely \textit{without} CDS that contain the FPPs ``I'' (upper-case), ``me'', ``my'', ``mine'', and ``myself''. As shown in \cref{table:CD_Category_Prevalence}: PR$_1$, we find that their removal does not significantly alter the observed effect. The respective confidence intervals resulting from our removal of FPP schemata change, but most overlap with those obtained from an analysis that includes the full set of CDS (see \cref{table:CD_Category_Prevalence}: PR$_A$ vs \cref{table:CD_Category_Prevalence}: PR$_1$). This demonstrates that our observations are not a product of the presence of first-person pronouns in our set of CDS. Note that we could not determine any values for \emph{Personalizing} because its CDS all contain first-person pronouns (see Appendix Fig. 5). 

\subsection{Robustness to CDS changes}

To determine the sensitivity of our results to the particular choice of CDS, we re-calculated PR values between the Depressed and Random Sample cohorts, but instead of re-sampling our Depressed and Random Sample cohort, we randomly re-sampled (with replacement) the set of 241 CDS n-gram. The 95\% CI of the resulting distribution of PR values then indicates how sensitive our results are to random changes of our CDS set. The results of this analysis are shown in \cref{table:CD_Category_Prevalence}: PR$_C$. We observe slight changes in the dispersion of the resulting distribution of PR values, but the median values and 95\% CIs remain largely unchanged. As before, the 95\% CIs continue to exclude $1.000$ for all CD types, except \emph{Mindreading}, \emph{Should Statements}, \emph{Fortune-telling}, and \emph{Catastrophizing}, and we can continue to reject the null-hypothesis that PR values are similar between the Depressed and Random Sample cohort for nearly all CD types. Furthermore, as shown in \cref{table:CD_Category_Prevalence}, the 95\% CIs of PR$_C$ and PR$_A$ largely overlap across all CD types indicating our results are robust to random changes of our cohort samples as well as our CDS set.

\section{Discussion}
In an online sample of individuals, we emphloyed a theory-driven approach to measure linguistic markers that may indicate cognitive vulnerability to depression, according to CBT theory. We defined a set of Cognitive Distortion Schemata (CDS) that we grouped along 12 widely accepted types of distorted thinking and compared their prevalence between two cohorts of Twitter users: one of individuals who self-identified as having received a clinical diagnosis of depression and the other a similar random sample.

As hypothesized, the Depressed cohort use significantly more CDS of distorted thinking in their online language than the Random Sample, particularly schemata associated with \emph{Personalizing} and \emph{Emotional Reasoning}. We observed significantly elevated levels of CDS across nearly all CD types, sometimes more than twice as much, but did not find a statistically significant elevated prevalance among the Depressed cohort for two specific types, namely \emph{Fortune-telling} and \emph{Catastrophizing}. This may be due to the difficulty of capturing these specific cognitive distortions in the form of a set of 1 to 5-grams as their expression in language can involve an interactive process of conversation and interpretation. Of note, our findings are not explained by the use of first-person pronouns or more negatively loaded language, both of which had been identified in past research as markers of depressed individuals. These results shed a light on how depression may affect public discourse on social media, but also reveals the degree to which depressogenic language is manifested in the colloquial language of social media platforms. This is of social relevance given that these platforms are specifically designed to propagate information through the social ties that connect individuals on a global scale.

An advantage of studying theory-driven differences between the language of depressed and non-depressed individuals, as opposed to a purely data-driven or machine learning approach, is that we can explicitly use the principles underpinning CBT to understand the cognitive and lexical components that may shape depression. Cognitive behavioral therapists have developed a set of strategies to challenge the distorted thinking that is characteristic of depression. Preliminary findings suggest that specific language can be related to specific therapeutic practices and seems to be related to outcomes\cite{ewbank2019quantifying}. These practices, however, have largely been shaped by a clinical understanding and not necessarily informed by objective measures of how patterns of language can determine the path of recovery. 

Our results suggest a path for mitigation and intervention, including applications that engage individuals suffering from mood disorders such as major depressive disorder via social media platforms and that challenge particular expressions and types of depressogenic language. Future characterizations of the relations between depressogenic language and mood may aid in the development of automated interventions (e.g., ``chatbots'') or suggest promising targets for psychotherapy. Another approach that has shown promise in leveraging social media for the treatment of mental health problems involves ``crowdsourcing'' the responses to cognitively-distorted content\cite{morris2015efficacy}.

Several limitations of our theory-driven approach should be considered. First, we rely on self-reported depression diagnoses on social media which have not been independently verified by a clinician. However, the potential inaccuracy of this inclusion criterion would reduce the observed effect sizes (PR values between cohorts) due to the larger heterogeneity of our cohorts. Consequently, our results are likely not an artifact of the accuracy of our inclusion criterion. Second, our lexicon of CDS was composed and approved by a panel of 9 experts who may have been only partially successful in capturing all n-grams used to express distorted ways of thinking. Nevertheless, a significant portion of CDS in our set did not occur in our collections of Twitter content, indicating the scope of our lexicon exceeds that of common online language. On a related note, the use of CDS n-grams implies that we measure distorted thinking by proxy, namely via language, and our observations may be therefore be affected by linguistic and cultural factors. Common idiosyncratic or idiomatic expressions may syntactically represent a distorted form of thinking, but no longer do in practice. For example, an expression such as ``literally the worst'' may be commonly emphloyed to express dismay, without necessarily involving the speaker experiencing a distorted mode of thinking.  Third, both cohorts were sampled from Twitter, a leading social media platform, whose use may be associated with higher levels of psychopathology and reduced well-being\cite{lin2016association,keles:systematic2019,kelly:social2019}. We may thus be observing elevated or biased rates of distorted thinking in both cohorts as a result of platform effects. However, we report relative prevalence numbers with respect to a carefully construed random sample, which likely compensates for this effect. Furthermore, recent analysis indicates that representative samples with respect to psychological phenomena can be obtained from social media content\cite{traditional:cattuto2020}. This is an important discussion in computational social science that will continue to be investigated. Data-driven approaches that analyze natural language in real-time will continue to complement theory-driven work such as ours. 

\section{Materials and Methods}
\subsection{Data and sample construction} Using the Twitter Application Program Interface (API) and the IUNI OSoMe\cite{davis2016osome} (a service which provides searchable access to the Twitter ``Gardenhose'', a 10\% sample of all daily tweets), we search for tweets that matched both ``diagnos*'' and ``depress*.'' The resulting set of tweets are then filtered for matching the expressions ``i'', ``diagnos*'', ``depres*'' in that order in a case-insensitive manner allowing insertions to match the greatest variety of diagnosis statements, e.g.~a tweet that states ``\emph{I} was in fact just \emph{diagnos}ed with clinical \emph{depress}ion'' would match. Finally, to ensure we are only including true self-referential statements of a depression diagnosis, a team of 3 experts manually removed quotes, jokes, and external references. For each qualifying diagnosis tweet we retrieve the timeline of the corresponding Twitter user using the Twitter \emph{user\_timeline} API endpoint~\footnote{\url{ https://developer.twitter.com/en/docs/tweets/timelines/api-reference/get-statuses-user_timeline}}. Subsequently, we remove all non-English tweets  (Twitter API machine-detected``lang'' field), all retweets, and tweets that contain ``diagnos*'' or ``depress*'', but not a valid diagnosis statement. The resulting Depressed cohort contains 1,207 individuals and 1,759,644 tweets ranging from from May 2008 to September 2018.

To compare CDS prevalence rates of the Depressed cohort to a baseline, we construct a Random Sample cohort of individuals. To do so, we collect a large sample of random tweets in 3 weeks (i.e. September 1-8, 2017, March 1-8, 2018, and September 1-8, 2018) from the IUNI OSOME\cite{davis2016osome}. We extract all Twitter user identifiers from these tweets (N=588,356), and retain only those that specified their geographical location and were not already included in our Depressed cohort. To equalize platform, interface, and behavioral changes over time, we select a sub-sample of these individuals such that the distribution of their account creation dates matches those of the Depressed cohort, resulting in an initial set of 9,525 random individuals. Finally, we harvested the Twitter timelines of these users and filtered the obtained data in the same way as described for the Depressed cohort. Since some user data was found to be no longer publicly available and others have no tweets left after our filters, our final Random Sample Cohort consists of 8,791 individuals and a total 8,498,574 tweets.

The code and data used in this analysis are freely available at \url{https://github.com/kbathina/CDs_Depressed_Twitter}. Upon reasonable request we will provide all Twitter user IDs and tweet IDs to reproduce our results.

\subsection{Prevalence Ratios}
For each Twitter user $u$ in our sample, we retrieved a timeline $T_u$ of their time-ordered $k$ most recent tweets, $T_u=\{t_1, t_2, \cdots, t_k\}$. We also defined a set $C = \{c_1, c_2, \cdots, c_N\}$ of n-grams where $N=241$ (see \cref{table:CD_classes}) with varying $n \in[1,5]$ number of terms. The elements of set C are intended to represent the lexical building blocks of expressing cognitive distortions (see \cref{table:CD_classes} and Appendix Table 7). We introduce a CDS matching function $\mathcal{F}_C(t) \rightarrow \{0,1\}$, which maps each individual tweet $t$ to either $0$ or $1$ according to whether a tweet $t$ contains one or more of the schemata in set $C$. Note that the range of $\mathcal{F}_C(t)$ is binary, thus a tweet that contains more than one CDS still counts as $1$.

The \textit{within-subject} prevalence of tweets for individual $u$ is defined as the ratio of tweets that contain a CDS in $C$ over all tweets in their timeline $T_u$:
\[
P_C(u) = \frac{\sum_{t \in T_u} \mathcal{F}_C(t)}{\left\vert T_u\right\vert}
\]

Our sample is separated into two cohorts: one of 1,207 Depressed and another of 8,791 Random Sample individuals. We denote the set of all individuals in the depressed cohort $D = \{u_1, u_2, \cdots, u_{1207}\}$ and random sample cohort $R = \{u_1, u_2, \cdots, u_{8791}\}$. Hence, the sets of all tweets written by users in the Depressed and Random Sample cohorts are defined as:

\begin{eqnarray}
T_D = {\textstyle\bigcup\limits_{u \in D}} T_u & \textrm{and} & 
T_R = {\textstyle\bigcup\limits_{u \in R}} T_u
\end{eqnarray}

We can then define the Prevalence ($P$) of tweets with CDS $C$ for each the Depressed ($D$) and Random Sample ($RS$) cohorts as follows:

\begin{eqnarray}
P_C(D) = \frac{\sum_{t \in T_D} \mathcal{F}_C(t)}{\left\vert T_D \right\vert} & & 
P_C(R) = \frac{\sum_{t \in T_R} \mathcal{F}_C(t)}{\left\vert T_R \right\vert}
\end{eqnarray}

or, informally, the ratio of tweets that contain any CDS over all tweets written by the individuals of that cohort.

Consequently, the Prevalence Ratio ($PR$) of CDS in set $C$ between the two cohorts $D$ and $R$, denoted $PR_C(D,R)$, is defined simply as the ratio of their respective CDS prevalence $P_C(T_D)$ and $P_C(T_R)$ in the tweet sets $T_D$ and $T_r$ respectively:

\begin{equation}
PR_c(D,R) = \frac{P_C(D)}{P_C(R)}
\end{equation}

If $PR_C(D,R) \simeq 1$ the prevalence of CDS in the tweets of the depression cohort are comparable to their prevalence in the tweets of the random sample. However, any value $PR_C(D,R) \ll1$ or $PR_C(D,R) \gg 1$ may indicate a significantly higher prevalence in each respective cohort. Here we use $\gg 1$ and $\ll 1$ to signifiy that a PR value is significantly higher or lower than 1 respectively, which we asses by whether its 95\% CI includes $1$ or not (see Bootstrapping below).

\subsection{Bootstrapping estimates} The estimated Prevalence and Prevalence Ratio can vary with the particular composition of either our set $C$ (CDS n-grams) or the set of individuals in our Depressed and Random Sample cohorts, respectively $D$ and $R$. We verify the reliability of our results by randomly re-sampling either $C$ or both $D$ and $R$, with replacement. This is repeated $B = 10000$ number of times, leading to a set of re-sampled CD sets or cohort samples. Each of these $B$ number of re-samples of either (1) the set of CDS $C$ or (2) or the sets $D$ and $C$ of all individuals in our Depressed and Random Sample cohorts results in $B$ number of corresponding Prevalence or Prevalence Ratio values: 
\begin{eqnarray}
P^* = \{P_1^*, P_2^*, \cdots, P_B^*\}, & PR^* = \{PR_1^*, PR_2^*, \cdots, PR_B^*\}
\end{eqnarray}
The distributions of $P^*$ and  $PR^*$ are then characterized by their median ($\mu_{50}$) and their 95\% confidence interval ($[\mu_{2.5}, \mu_{97.5}]$). A 95\% confidence interval of a PR that does not contain $1$ is held to indicate a significant difference in prevalence between the two cohorts.

\section{Acknowledgements}
We thank Luis M. Rocha for his feedback on the general methodology and terminology, as well as Drs. Keith Dobson, Rob DeRubeis, Christian Webb, Stephan Hoffman, Nikolaos Kazantzis, Judy Garber, and Robin Jarrett for their feedback on the content of our list of CDS. Johan Bollen thanks NSF grant \#SMA/SME1636636, the Indiana University ``Grand Challenges - Prepared for Environmental Change" PR-IUB grant, Wageningen University, and the ISI Foundation for their support.

\section*{References}
\bibliographystyle{naturemag}
\bibliography{CDratio}

\section*{Appendix}

\begin{table*}[!htbp]
\scriptsize
\begin{center}
\begin{tabular}{lcc}
 & $P_C(D)$ (\%) & $P_C(R)$ (S\%)  \\\cline{1-3}
All CDs        & 21.838 & 18.407 \\ \cline{1-3}
Dichotomous Reasoning & 16.650 & 13.933  \\
Should statements  & 3.191 & 2.896  \\
Magnification and Minimization & 1.992 & 1.851 \\
Labeling and mislabeling  & 1.199 & 0.903   \\
Mindreading & 1.168 & 1.026  \\
Personalizing   & 1.026 & 0.427 \\
Overgeneralizing  & 0.752 & 0.476  \\
Disqualifying the Positive &  0.081 & 0.060 \\        
Emotional Reasoning  & 0.053 & 0.023 \\ 
Fortune-telling  & 0.047 & 0.050 \\            
Mental Filtering  & 0.024 & 0.016 \\
Catastrophizing  & 0.014 & 0.019 \\    \hline    
\end{tabular}
\end{center}
\caption{Raw CDS Prevalence in tweets from the Depressed and Random Sample cohorts in decreasing rank order.}
\end{table*}

\begin{figure*}[!htbp]
\centering
\includegraphics[width=.8\linewidth]{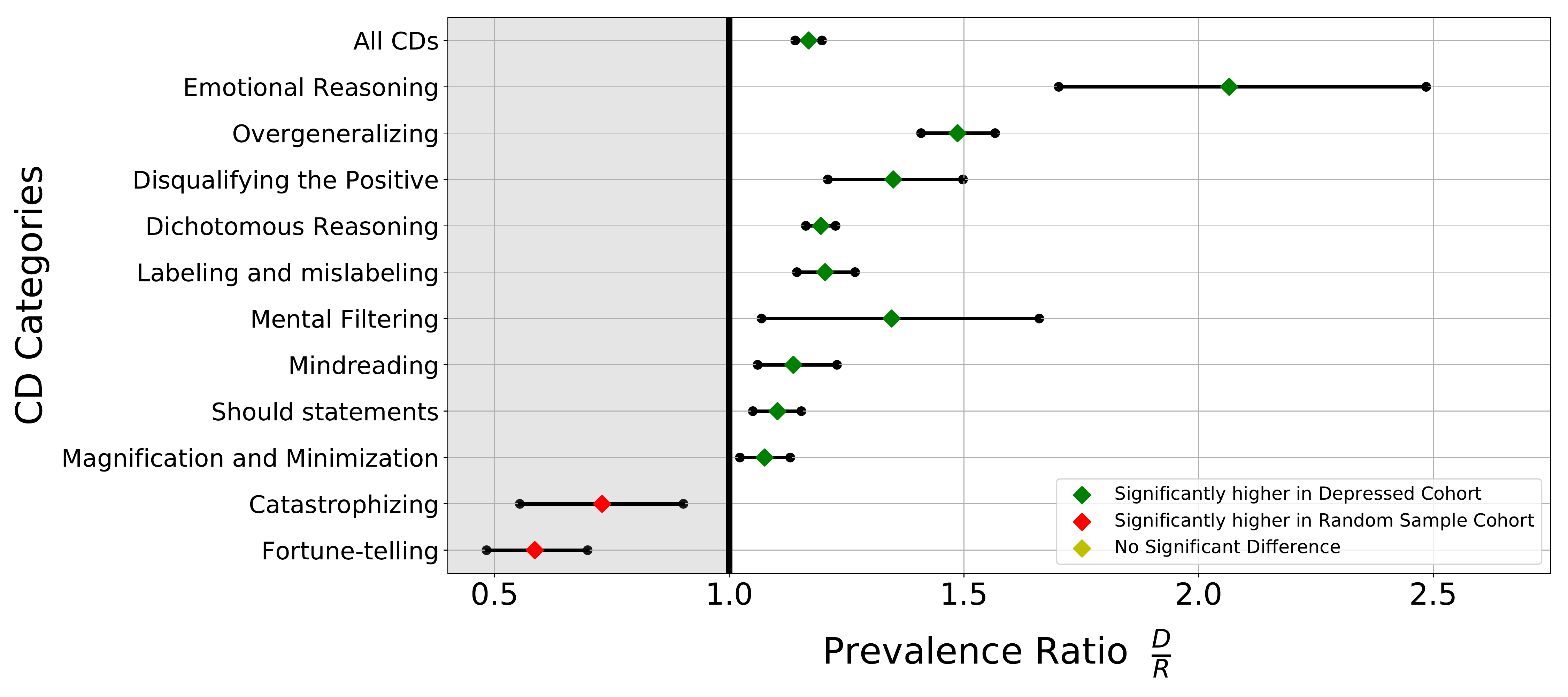}
\caption{Prevalence Ratio of CD markers \textit{without First-Person Pronouns (FPP)} for each type of cognitive distortion between the Depressed and Random Sample cohort. The two black dots represent the 2.5th and 97.5th percentile of the bootstrapped distribution of means. The diamonds indicate the location of the median, with green, yellow, and red diamonds indicating Prevalence Ratio values significantly higher than parity (1) for the Depressed cohort, no significant difference from parity (yellow), and sigfificantly lower prevalence for the Depressed cohort than the Random Sample (red). \emph{Catastrophizing} and \emph{Fortune-telling} are significantly more prevalent in the Random Sample population ($\ll1$) while the other types of CDS are more significant in the Depressed cohort ($\gg1$).}
\end{figure*}

\begin{table*}[!htbp]
\begin{center}
\scriptsize
\begin{tabular}{p{1cm}p{4.5cm}p{0.8cm}C{1.8cm}|p{0.8cm}C{1.8cm}|p{0.8cm}C{1.8cm}}
& &  \multicolumn{2}{c}{$PD_A$} 
& \multicolumn{2}{c}{$PD_1$} & \multicolumn{2}{c}{$PD_C$} \\
& & median & 95\% CI & median & 95\% CI & median & 95\% CI \\\cline{2-8}
\multirow{11}{*}{$P_D \gg P_R$}    & 
All CDs                                           & 3.431$^{*}$  & [2.912, 3.939]   & 3.069$^{*}$   & [2.568, 3.561]   & 0.021$^{*}$   & [0.013, 0.031]\\ \cline{2-8}
& Dichotomous Reasoning                           & 2.714$^{*}$  & [2.293, 3.130]   & 2.714$^{*}$   & [2.294, 3.130]   & 0.143$^{*}$   & [0.072, 0.231]\\
& Personalizing                                   & 0.599$^{*}$  & [0.537, 0.663]   & /             & /                & 0.041$^{*}$   & [0.003, 0.110]\\
& Overgeneralizing                                & 0.276$^{*}$  & [0.242, 0.311]   & 0.177$^{*}$   & [0.151, 0.203]   & 0.013$^{*}$   & [0.003, 0.025]\\
& Labeling and Mislabeling                        & 0.296$^{*}$  & [0.244, 0.349]   & 0.146$^{*}$   & [0.104, 0.189]   & 0.007$^{*}$   & [0.002, 0.013]\\
& Should Statements                               & 0.297$^{*}$  & [0.149, 0.436]   & 0.297$^{*}$   & [0.149, 0.436]   & 0.060         & [-0.050, 0.213]\\
& Mindreading                                     & 0.140$^{*}$  & [0.063, 0.233]   & 0.140$^{*}$   & [0.063, 0.233]   & 0.002         & [-0.001, 0.007]\\
& Magnification and Minimization                  & 0.139$^{*}$  & [0.043, 0.235]   & 0.139$^{*}$   & [0.043, 0.235]   & 0.018$^{*}$   & [0.004, 0.033]\\
& Emotional Reasoning                             & 0.030$^{*}$  & [0.025, 0.036]   & 0.006$^{*}$   & [0.004, 0.008]   & 0.004$^{*}$   & [0.001, 0.009]\\     
& Disqualifying the Positive                      & 0.021$^{*}$  & [0.013, 0.029]   & 0.021$^{*}$   & [0.013, 0.029]   & 0.001$^{*}$   & [0.000, 0.003]\\        
& Mental Filtering                                & 0.008$^{*}$  & [0.005, 0.010]   & 0.002$^{*}$   & [0.000, 0.004]   & 0.001$^{*}$   & [0.000, 0.001] \\\cline{2-8}
\multirow{2}{*}{$P_D \ll P_R$} & Fortune-telling  & -0.002       & [-0.009, 0.003]  & -0.012        & [-0.017, -0.008] & -0.000        & [-0.002, 0.003]\\    
& Catastrophizing                                 & -0.005       & [-0.010, -0.002] & -0.005        & [-0.010, -0.002] & -0.000        & [-0.001, 0.000]\\    
\end{tabular}
\end{center}
\caption{Prevalence Difference (PD) percentages for set of CDS $C$, defined as $(P_C(D) - P_C(R)) \times 100$, and its 95\% CIs between the Depressed and Random Sample cohort. PD values $\gg 0$ indicate a significantly higher prevalence in the Depressed sample (marked by $^*$). Values are calculated under 3 distinct conditions, labeled $PD_A$: PD percentage difference values for the entire set of CD markers, $PD_1$: PD percentage difference values for CD markers without first-person pronouns (``I'',``me'', ``my'', ``mine'', and ``myself''), and $PD_C$: PD percentage values with a 95\% CIs resulting from re-sampling the CD marker set 10,000 times, instead of the individuals in the Depressed and Random Sample cohorts (see Main: Materials and Methods).}
\end{table*}

\begin{figure*}[!htbp]
\centering
\includegraphics[width=1.0\linewidth]{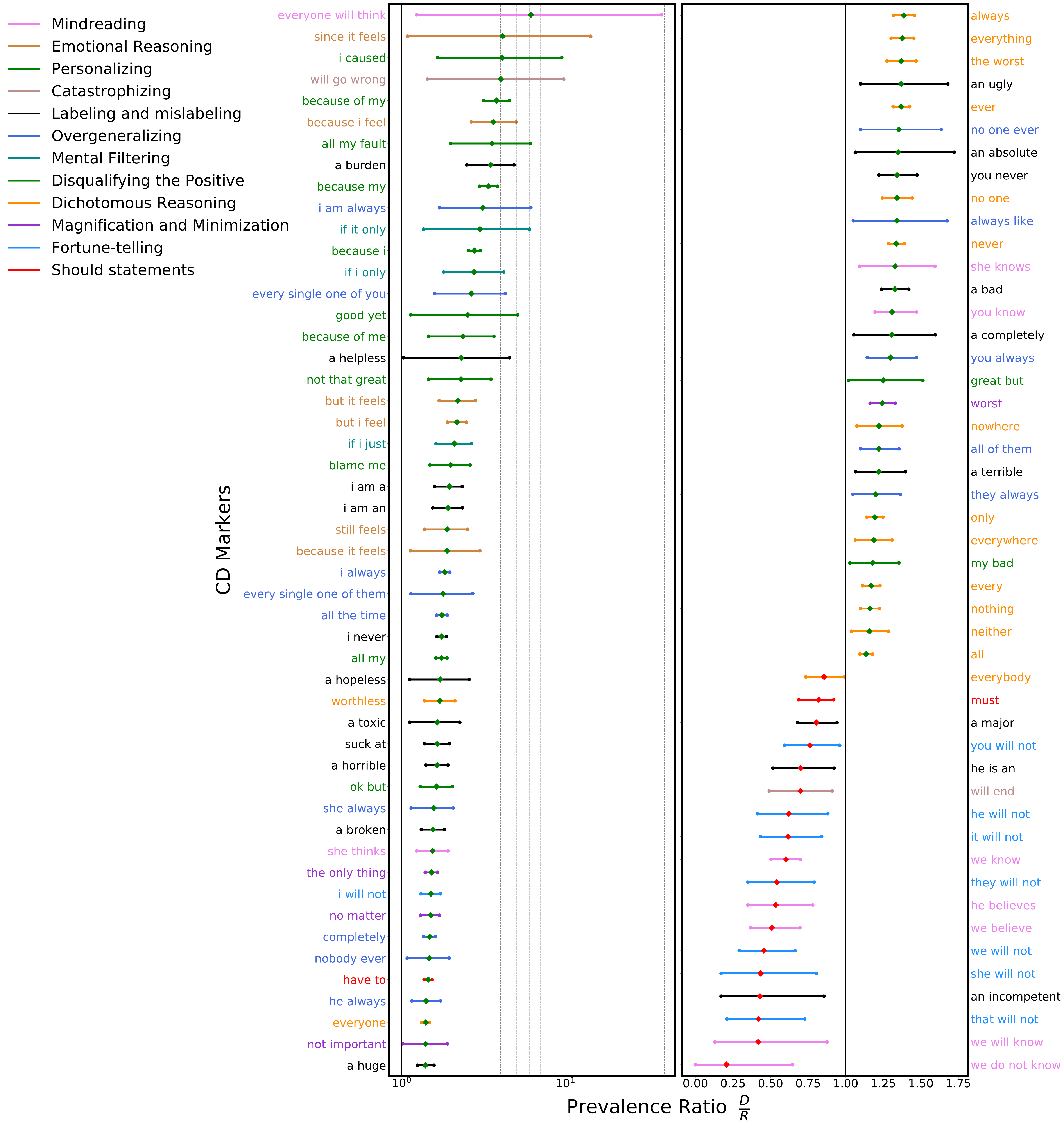}
\caption{Median and 95\% confidence interval of the bootstrapped PR means of CDS n-grams that were significantly more prevalent, i.e. $PR \ll 1$ or $PR \gg 1$, in the Depressed or Random Sample cohort. CDS n-grams are colored by the type of CD they belong to.}
\end{figure*}  

\begin{table*}[!htbp]
\scriptsize
\begin{center}
\begin{tabular}{p{4.5cm}p{10cm}}
Category                            &   CD Markers        \\ \hline

\emph{Catastrophizing}            &   \textit{will fail, will go wrong, will end, will be impossible, will not happen, will be terrible, will be horrible, will be a catastrophe, will be a disaster, will never end, will not end}\\

\emph{Dichotomous Reasoning}      &   \textit{only, every, everyone, everybody, everything, everywhere, always, perfect, the best, all, not a single, no one, nobody, nothing, nowhere, never, worthless, the worst, neither, nor, either or, black or white, ever}\\
                                    
\emph{Disqualifying the Positive} &   \textit{great but, good but, OK but, not that great, not that good, it was not, not all that, fine but, acceptable but, great yet, good yet, OK yet, fine yet, acceptable yet
}\\
                                    
\emph{Emotional Reasoning}        &   \textit{but I feel, since I feel, because I feel, but it feels, since it feels, because it feels, still feels
}\\
                                    
\emph{Fortune-telling}            &   \textit{I will not, we will not , you will not, they will not, it will not, that will not, he will not, she will not
}\\

\emph{Labeling and Mislabeling}   &   \textit{I am a, he is a, she is a, they are a, it is a, that is a, sucks at, suck at, I never, he never, she never, you never, we never, they never, I am an, he is an, she is an, they are an, it is an, that is an, a burden, a complete, a completely, a huge, a loser, a major, a total, a totally, a weak, an absolute, an utter, a bad, a broken, a damaged, a helpless, a hopeless, an incompetent, a toxic, an ugly, an undesirable, an unlovable, a worthless, a horrible, a terrible}\\

\emph{Magnification and Minimization} &   \textit{worst, best, not important, not count, not matter, no matter, the only thing, the one thing
}\\
                                        
\emph{Mental Filtering}           &   \textit{I see only, all I see, all I can see, can only think, nothing good, nothing right, completely bad, completely wrong, only the bad, only the worst, if I just, if I only, if it just, if it only
}\\
                                    
\emph{Mindreading}                &   \textit{everyone believes, everyone knows, everyone thinks, everyone will believe, everyone will know, everyone will think, nobody believes, nobody knows, nobody thinks, nobody will believe, nobody will know, nobody will think, he believes, he knows, he thinks, he does not believe, he does not know, he does not think, he will believe, he will know, he will think, he will not believe, he will not know, he will not think, she believes, she knows, she thinks, she does not believe, she does not know, she does not think, she will believe, she will know, she will think, she will not believe, she will not know, she will not think, they believe, they know, they think, they do not believe, they do not know, they do not think, they will believe, they will know, they will think, they will not believe, they will not know, they will not think, we believe, we know, we think, we do not believe, we do not know, we do not think, we will believe, we will know, we will think, we will not believe, we will not know, we will not think, you believe, you know, you think, you do not believe, you do not know, you do not think, you will believe, you will know, you will think, you will not believe, you will not know, you will not think}
\\
                                    
\emph{Overgeneralizing}           &   \textit{all of the time, all of them, all the time, always happens, always like, happens every time, completely, no one ever, nobody ever, every single one of them, every single one of you, I always, you always, he always, she always, they always, I am always, you are always, he is always, she is always, they are always}\\
                                    
\emph{Personalizing}              &   \textit{all me, all my, because I, because my, because of my, because of me, I am responsible, blame me, I caused, I feel responsible, all my doing, all my fault, my bad, my responsibility}\\
                                    
\emph{Should Statements}          &   \textit{should, ought, must, have to, has to}\\
                                    
\end{tabular}
\end{center}
\caption{Cognitive Distortion types and corresponding schemata}
\end{table*}

\begin{figure*}[!htbp]
\centering
\includegraphics[width=.8\linewidth]{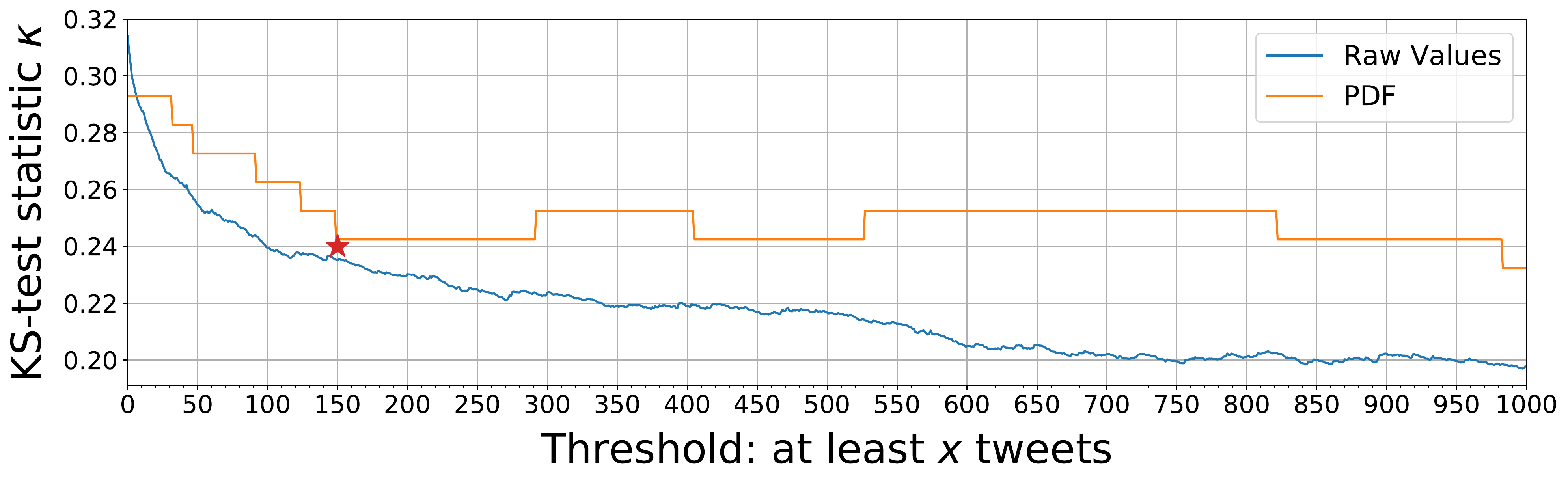}
\caption{
To prevent the calculation of within-subject prevalence values for individuals with too few data points, we calculate the within-subject prevalence values only for individuals with a given minimum number of tweets (150). To determine the sensitivity of our results to the chosen value of this threshold, we calculate the KS-test statistic of the distributions of within-subject CDS prevalence given each threshold value (x-axis). The blue line shows the KS statistic for the raw prevalence distribution while the orange line shows the KS statistic for the Probability Distribution Function (PDF). We visually determine an inflection point in both graphs at 150 minimum tweets at which point the corresponding p-value for the KS-test stabilizes at $2.8 * 10^{-45}$ and $5.8 * 10^{-3}$ respectively. Consequently we calculate within-subject prevalence values only for individuals in either cohort for which we retrieved at least 150 tweets. We can reject the null-hypothesis that the distributions were drawn from the same population for \textit{all} threshold values indicating the conclusions we draw from a comparison of the within-subject CDS prevalences between the Depressed and the Random Sample cohort are not sensitive to our choice of threshold. It is also important to note that we do not apply this individual threshold to the calculation of between-group Prevalence Ratios, since this indicator is calculated at the group level pertaining to millions of tweets and not calculated at the level of individual users.}
\end{figure*}

\end{document}